# Estimating required 'lockdown' cycles before immunity to SARS-CoV-2: Model-based analyses of susceptible population sizes, 'S0', in seven European countries including the UK and Ireland

Rosalyn J. Moran[1*], Erik D. Fagerholm[1], Maell Cullen[1], Jean Daunizeau[2], Mark P. Richardson[3], Steven Williams[1], Federico Turkheimer[1], Rob Leech[1], Karl J. Friston[4]

[1.] Department of Neuroimaging, Institute of Psychiatry, Psychology & Neuroscience, King's College London, De Crespigny Park, London, SE5 8AF, UK.

[2.] ICM, INSERM UMRS 1127, Paris, France

[3.] Department of Basic and Clinical Neuroscience, Institute of Psychiatry, Psychology & Neuroscience, King's College London, De Crespigny Park, London, SE5 8AF, UK.

[4.] Wellcome Trust Centre for Human Neuroimaging, Institute of Neurology, University College London, 12 Queen Square, London WC1N 3BG, UK.

**\*Address for Correspondence**
Rosalyn.moran@kcl.ac.uk
**Academic Neurosciences Centre**
**Centre for Neuroimaging Sciences**
**De Crespigny Park**
**London**
**SE5 8AF**
**+44 207 848 6173**

**Abstract**

**Background:** Following stringent social distancing measures, some European countries are beginning to report a slowed or negative rate of growth of daily case numbers testing positive for the novel coronavirus. The notion that the first wave of infection is close to its peak begs the question of whether future peaks or 'second waves' are likely. We sought to determine the current size of the effective (i.e. susceptible) population for seven European countries—to estimate immunity levels following this first wave. We compare these numbers to the total population sizes of these countries, in order to investigate the potential for future peaks.

**Methods:** We used Bayesian model inversion to estimate epidemic parameters from the reported case and death rates from seven countries using data from late January 2020 to April 5$^{th}$ 2020. Two distinct generative model types were employed: first a continuous time dynamical-systems implementation of a Susceptible-Exposed-Infectious-Recovered (SEIR) model and second: a partially observable Markov Decision Process (MDP) or hidden Markov model (HMM) implementation of an SEIR model. Both models parameterise the size of the initial susceptible population ('S0'), as well as epidemic parameters. Parameter estimation ('data fitting') was performed using a standard Bayesian scheme (variational Laplace) designed to allow for latent unobservable states and uncertainty in model parameters.

**Results:** Both models recapitulated the dynamics of transmissions and disease as given by case and death rates. The peaks of the current waves were predicted to be in the past for four countries (Italy, Spain, Germany and Switzerland) and to emerge in 0.5 – 2 weeks in Ireland and 1-3 weeks in the UK. For France one model estimated the peak within the past week and the other in the future in two weeks. Crucially, *Maximum a posteriori* (MAP) estimates of S0 for each country indicated effective population sizes of below 20% (of total population size), under both the continuous time and HMM models. Using for all countries—with a Bayesian weighted average across all seven countries and both models, we estimated that 6.4% of the total population would be immune. From the two models the maximum percentage of the effective population was estimated at 19.6% of the total population for the UK, 16.7% for Ireland, 11.4% for Italy, 12.8% for Spain, 18.8% for France, 4.7% for Germany and 12.9% for Switzerland.

**Conclusion:** Our results indicate that after the current wave, a large proportion of the total population will remain without immunity. This suggests that in the absence of strong seasonal effects, new medications or more comprehensive contact tracing, a further set of epidemic waves in different geographic centres are likely. These findings may have implications for 'exit strategies' from any lockdown stage.

**Introduction**

As of early April 2020, the Coronavirus pandemic has reached different epidemic stages across the world. France was the earliest affected country in Europe with its first reported cases on 24th January 2020 (Reusken, Broberg et al. 2020) with cases reported shortly after in Germany, then the United Kingdom, Italy, Spain, Switzerland and in Ireland on February 29th. Subsequently outbreaks have emerged across the European continent. The daily rates of new confirmed cases of the Covid-19 virus (SARS-CoV-2) have begun to decrease in some of these countries; in particular, in Italy and Spain, with promising signs that extensive social distancing measures have been effective and that these countries have reached or are past 'the peak' of infections. Epidemiological models that predict the progression of populations from Susceptible (S) to Exposed (E), Infected (I) and Recovered (R) (SEIR models (Kermack and McKendrick 1927)) can be used to investigate the properties of these peaks, given the initial susceptibility of a population. For SARS-CoV-2, no (or limited) immunity can be assumed a priori in humans and thus the majority of the entire population is deemed susceptible (Team 2020).

Several studies have developed and simulated SEIR models using epidemic parameters to 'nowcast' and forecast transmission (Wu, Leung et al. 2020). Parameters of the model are being continuously improved and modified, such as reduced serial interval estimates (Nishiura, Linton et al. 2020, Yang, Zeng et al. 2020), initially derived from observed cases in the initial outbreak in Wuhan, China (Sun, Chen et al. 2020, Wang, Liu et al. 2020, Wang, Wang et al. 2020). In most studies, these compartmental models are applied as dynamic generative (i.e., causal or mechanistic) models that assume a set of parameters and predict cases or clinical resources (Moghadas, Shoukat et al. 2020) and intervention effects (Prem, Liu et al. 2020, Wells, Sah et al. 2020). The (initial) susceptible size of a population (termed 'S0') is assumed to be the size of a particular city, e.g. 10 million in Wuhan (Prem, Liu et al. 2020) or—for a country—is assumed to comprise of multiples of smaller city sized outbreaks e.g. 100k (Ferguson, Laydon et al. 2020). Such models have lent important insight into the likely disease and clinical trajectories of countries as a whole, enabling planning and management for predicted numbers of cases requiring hospitalization and ventilation (Moghadas, Shoukat et al. 2020).

Given the lockdowns around Europe which likely averted larger case surges (Wang, Liu et al. 2020), we sought to investigate the current effective population size in seven European countries. Therefore, we used the SEIR model to determine the initial population size (S0) that was susceptible (i.e., would eventually become infected) at the beginning of the first wave and thus determine the levels of immunity that might exist in these countries after this wave, (by assuming the susceptible population will eventually become infected and develop immunity). One approach, to perform this inverse modelling, is to apply dynamic causal modelling (Friston, Mattout et al. 2007)—enabling the incorporation of prior values for parameters (e.g. the serial interval, incubation period or number of daily contacts) and prior uncertainty about these values.

Quarantine and social isolation are often explicitly accounted for in SEIR models (Wearing, Rohani et al. 2005, Feng 2007, Ridenhour, Kowalik et al. 2018) making them appropriate for the current Government advised social distancing. Importantly, two particular forms of this SEIR model (with social distancing) have recently been developed (Friston, Parr et al. 2020, Moghadas, Shoukat et al. 2020) that also account for deaths following hospitalization or treatment via ventilation within an Intensive Care Unit (ICU). Specifically, they account for a potential time lag between becoming infected and developing acute respiratory distress. This makes these models putatively 'fit for purpose', when using death as well as case reporting data to fit or invert the models to recover (posterior) parameter values—and estimate their uncertainty.

We aimed to apply these two models to data from seven countries: including Ireland, the United Kingdom, Italy, Spain, France, Germany and Switzerland. Our goal was to estimate S0. One model, (the 'ODE model'—see Methods and (Moghadas, Shoukat et al. 2020)) is based on a classical compartment model where a person in an epidemic can occupy only one compartment or 'state' and moves from state to state: from Susceptible to eventually (through intermediate states) either Recovery or Death. The other model—the 'HMM model' (Friston, Parr et al. 2020)—features several factors that change together; including where a person is located (out of the home vs. in the home, for example), as well as their infectious, testing and clinical status. We apply both models to daily case and death reports to assess whether there is convergence on estimates of the initially susceptible (i.e., effective) population sizes.

**Methods**

*Data*

Data from a repository for the 2019 Novel Coronavirus at John's Hopkins University Center for Systems Science and Engineering (JHU CSSE) were used (Dong, Du et al. 2020). Using these date-stamped entries of daily reported cases and reported deaths, we extracted seven timeseries pairs for the countries (including all territories) of Ireland, the United Kingdom, Italy, Spain, Germany, France and Switzerland. Data records from January 22$^{nd}$ to April 5$^{th}$, 2020 were modelled. For the ordinary differential equation (ODE) model, daily cases and daily accumulated deaths were fitted while for the HMM model, daily cases and daily deaths were fitted, corresponding to the state equations (see (Friston, Parr et al. 2020)).

*Models*

ODE Model: A dynamic transmission model comprising a set of 12 coupled ordinary differential equations was adapted from Moghadas et al (Friston, Parr et al. 2020). The original model included 12 states for four different age categories. We simplified the model structure by collapsing across age (see Appendix A). The 12 states or compartments in this simplified model (flow function, Figure 1) described susceptible (S) individuals who became infected with the disease through exposure (E) to other infected individuals. Infected individuals comprised three categories, an Asymptomatic or subclinical state (A), a symptomatic state who would not require hospitalization (InH) and a symptomatic state who would require hospitalization (IH). Each of these infected categories could also self-isolate – representing three more states defined by lower *a priori* contacts. People in states InH and A were assumed to recover, while those in states (IH) would transition to either hospitalized (H) or ICU states (ICU). From these states people would recover (R) or die (D), Figure 1. Time constants of the mode included the incubation period, recovery period, time to self-isolate, time from symptom onset to hospitalization, time from ICU admission to death, time from non-ICU admission to death, length of stay in ICU and length of hospital stay. Parameters controlling proportions that entered branching states (e.g. proportion of all hospitalised cases admitted to ICU were also included (see Appendix A for full parameter list) as well as the transmission rate and contact per day either within or without self-isolation. Parameters were equipped with *a priori* values and optimisation was performed on log scale factors to ensure positivity (Appendix A) of these proportions and rate constants. To link these ODEs to the observed data we employed an observer function which assumed a variable rate of case reporting for symptomatic (without requiring hospitalization) and asymptomatic individuals. A priori we assumed that only 1% of individuals infected who were asymptomatic received tests. We assumed that 20% of symptomatic cases who do not require hospitalization receive tests. And that 100% of infected individuals who are hospitalised receive tests. The levels of 1% and 20% testing were free parameters in our model. The 100% for hospitalised tests was fixed. Finally, 100% of deaths were assumed to be recorded. Finally, we placed a prior on the initial number of individuals in each state. A priori, we assume 100 individuals in infected states. We tested two alternatives for S0:

In the ODE model (Model: ODE) we initialised S0 to 1 Million x $\theta\_S0$ individuals, where $\theta\_S0$ = 1. This parameter would be optimised for each individual dataset and so could accommodate total sizes; e.g. if $\theta\_S0$ = 4.9 *a posteriori* then the total population of Ireland would be considered initially susceptible.

We also tested a 'cities'-based version of the ODE model (Model: ODE_City) that might recapitulate the death and case rates for each country. For this, we altered the observer function and imposed a prior of 1 Million x $\theta\_S0$ individuals, where $\theta\_S0$ = 1. Then we scaled the case and death rates by the population in millions (See Appendix A for equations). For this if we obtained $\theta\_N0$ = 1 *a posteriori* then the total initial susceptible population would also correspond to the total population of Ireland, but the epidemic dynamic would comprise 4.9 distinct outbreaks.

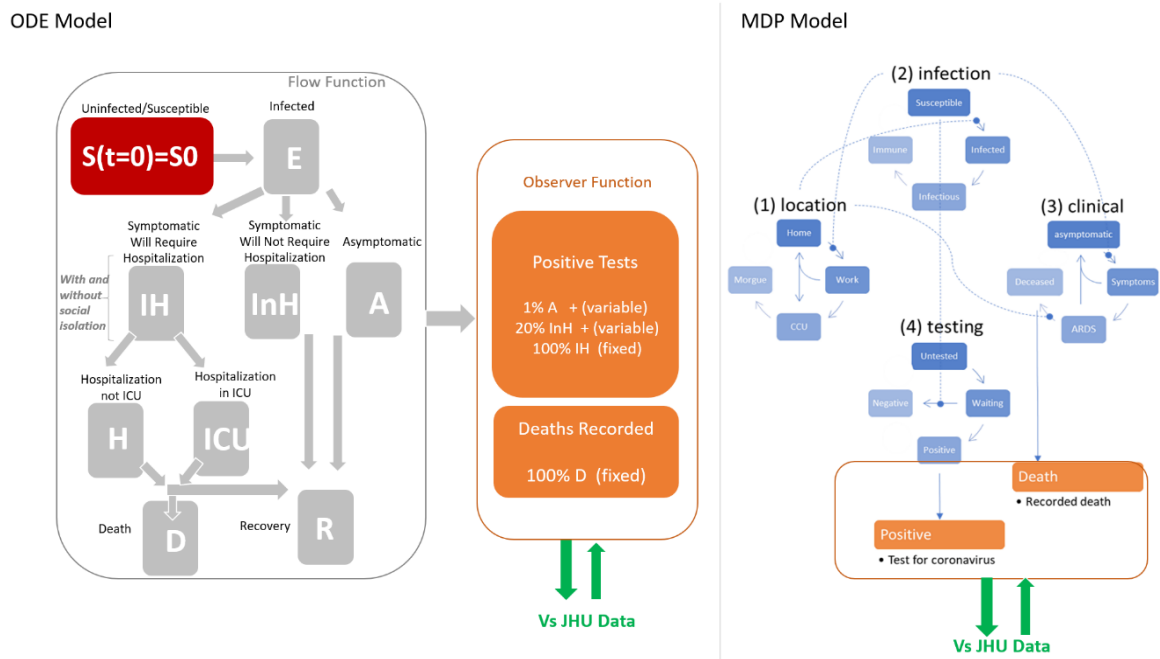

**Figure 1. Left: Flow function, or population dynamics. From the susceptible state (where initially at time t= 0S = S0) , the infected population will enter one of three categories: IH (infected requiring hospitalization), InH (infected NOT requiring Hospitalization), or A (asymptomatic). From the IH state, subjects transfer either to H (Hospital) or ICU, from which subjects transfer either to R or D. Both the InH and A states lead to Recovery (R). The observer function evaluates the dependent variable at each iteration of the integration process of the flow function. The resultant model-based data is compared against empirical JHU data. Right: Network showing the transition between states for the HMM model. The ODE model is a 12-compartmental model—with one factor with 12 states). The HMM model is a 256-compartment model—with four factors (location, infection and clinical) each with four levels, giving $4^4$ = 256 states. The structural difference between the ODE and HMM model rests upon the allowable combination of factors that describe the state of an individual in a population. For example, in the factorial (HMM) model it is possible to die from acute respiratory distress at (e.g. a care) home. Conversely, in the ODE model, one can only die after being in hospital. Certain transitions among these states are allowable. For example, in the 'testing' factor, an individual could transit from untested to waiting to a positive result. Or an individual could transition from untested to untested—where they remain untested even if they are infected (see Friston et al. 2020 for a priori probability values).**

HMM model: Our second model of the cases and death rates was a Dynamic Causal Model of Covid-19. See (Friston, Parr et al. 2020) for a complete description of the model. In brief, the model represents four factors describing location, infection status, test status and clinical status. Within each factor people may transition among four states probabilistically. The transitions generate predicted outcomes; for example, the number of people newly infected who have tested positive or the number of people newly infected who will remain untested. The location factor describes if an individual is at home, at work, in a critical care unit (CCU) or deceased. Similar to the early states in the ODE model, the HMM has a second factor describing infection status susceptible, infected, infectious or immune, where it is assumed that there is a progression from a state of susceptibility to immunity—through a period of (pre-contagious) infection to an infectious (contagious) status. The Clinical status factor comprises asymptomatic, symptomatic, acute respiratory distress syndrome (ARDS) or deceased. Finally, the fourth factor represents diagnostic status where an individual can be untested or waiting for the results of a test that can either be positive or negative. As with the ODE model, transitions amongst states are controlled by rate constants (inverse time constants) and non-negative probabilities. Similar to the ODE model above, we initialised (and set as priors) S0 to 1 Million x $\theta\_S0$ individuals, where $\theta\_S0$ = 1.

For the HMM and both ODE models (ODE model and ODE_City) to estimate the model parameters, we employed a standard (variational Laplace) Bayesian scheme to optimise parameters of corresponding DCM (spm_NLSI_GN) (Friston, Mattout et al. 2007).

**Results**

The key aim of our analysis was to estimate the likely immunity after the current set of cases and deaths. To ascertain the initial susceptibility S0, we examined the posterior estimate from both model types and its Bayesian credible intervals. However, first we examined the evidence for each model, relative to the worst performing model. We used two ODE models, with different constructs for epidemic sizes/meta-populations. The first ODE model (ODE) assumed a prior of 1 million susceptible individuals (S0), the. The second ODE model accounted for several effective populations of size 1 million (ODE_City). The third model was the HMM model, which also assumed a prior of 1 million initial susceptible individuals. Of all three models, ODE_City was the worst performing model for all countries data (Figure 2A).

From the two better performing models, we then estimated the effective population size of S0=S(t=0) as a proportion of the total population (Figure 2B). Taking a Bayesian average—across all models and countries—the estimated proportion of people that were initially susceptible at the start of this outbreak—and thus immune at the end of the outbreak—was 6.4% of the total population of each country.

The ODE model produced consistently higher estimates of S0 at the end of the wave than the HMM. These values suggest that after the current wave of cases, between 3 (lowest estimates for Ireland and the UK) and 12 (highest estimate for Germany) more cycles (with identical dynamics to those from Jan 22nd) would be required to bring the total population to probable herd immunity levels (we assume herd immunity of 60%, Figure 2B). We plot this fall in susceptibility state S (increase in immunity) over time, from the initial size S0 in figure 2C for the ODE and HMM models separately for Ireland and the UK (Figure 2C).

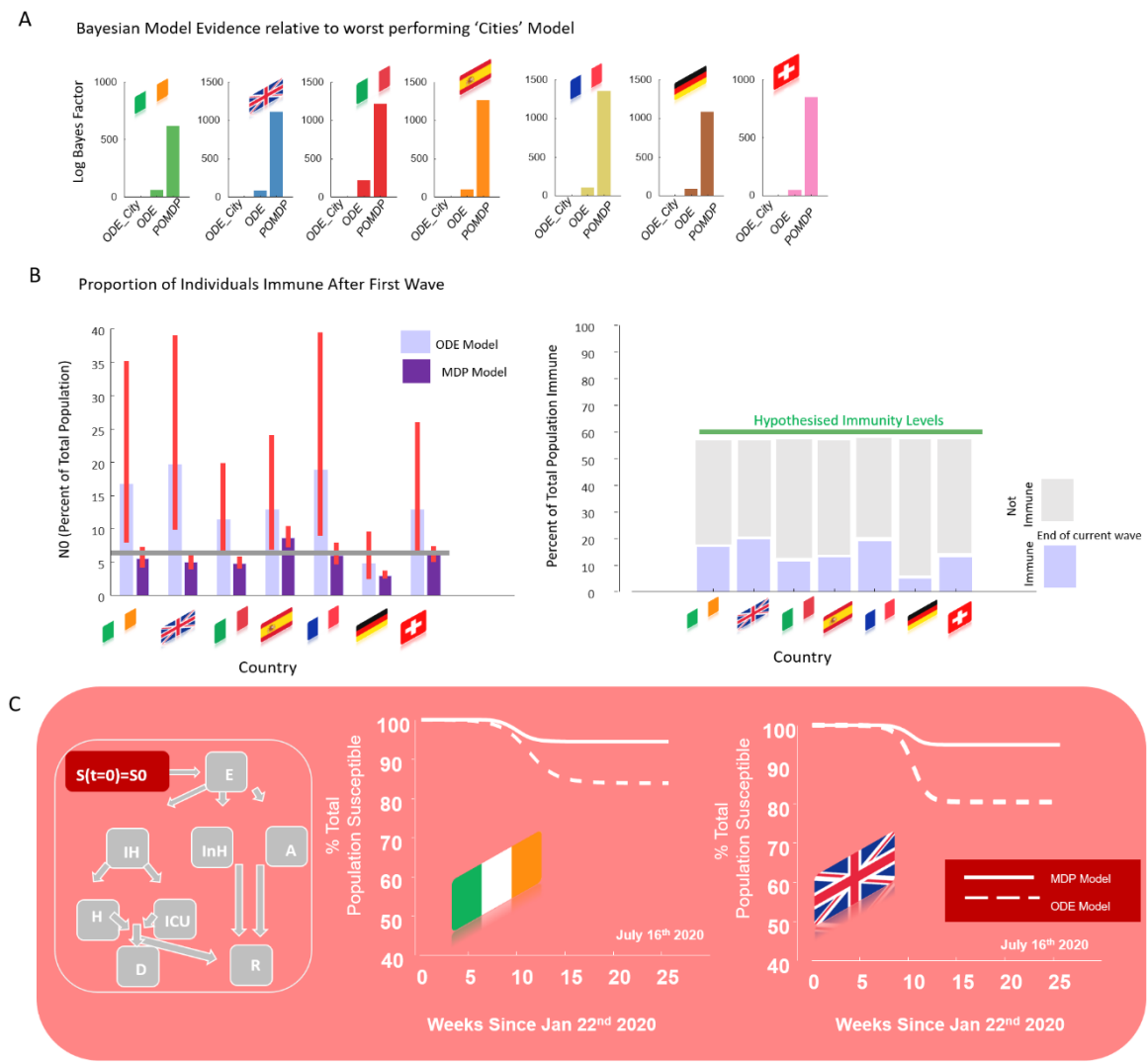

**Figure 2: A) lower bound on log model evidence given by variational free energy (Log Bayes Factor) for different models across countries. This shows that the models with a prior of 1 million S0 outperform the (multiple outbreak ODE_Cities) model, with priors of initial susceptible S0 equal to the total population of the country. B) From the two better performing models, we report the percent of the country's total population that are initially susceptible. We also plot the Bayesian parameter average of this percent, across all countries and models (6.4%). C) From the ODE model, we replot the fall in**

the susceptible population for Ireland and the UK as the dynamics of the current wave unfold. At the end of this wave large percentages of the population remain susceptible. The projections under the HHM are more pessimistic because the effective population size at the start of the first outbreak (i.e., susceptible population) is smaller (and more precise).

Our model inversion procedure produced fits to the data that recapitulated the rates up to April 5[th] for both models (Figure 3). Systematic differences in future predictions were observed however between the ODE and HMM models (though predictions were of similar orders of magnitude). For all countries the peak date and peak number of cases was higher for the ODE model. However, both models exhibited peaks for dates *in the past* for four countries (Italy, Spain, Germany and Switzerland). For France the models were discrepant with the ODE model predicted a peak in the future on April 20[th] and the HMM model estimating a peak had already occurred on April 7[th].

Peaks in the future were expected for Ireland and the UK. For Ireland, the peak reported case rate predictions were estimated at April 9[th] for the HMM and April 23[rd] for the ODE. The estimate of the number of daily cases at the peak were 720 cases and 392 cases for the ODE and HMM models. For the UK, the peak case rate predictions were estimated at April 11[th] for the HMM and April 17[th] for the ODE model. The peak case rates (i.e. tested cases) were estimated at 9304 daily cases for the ODE and 5411 daily cases for the HMM models (Figure 3).

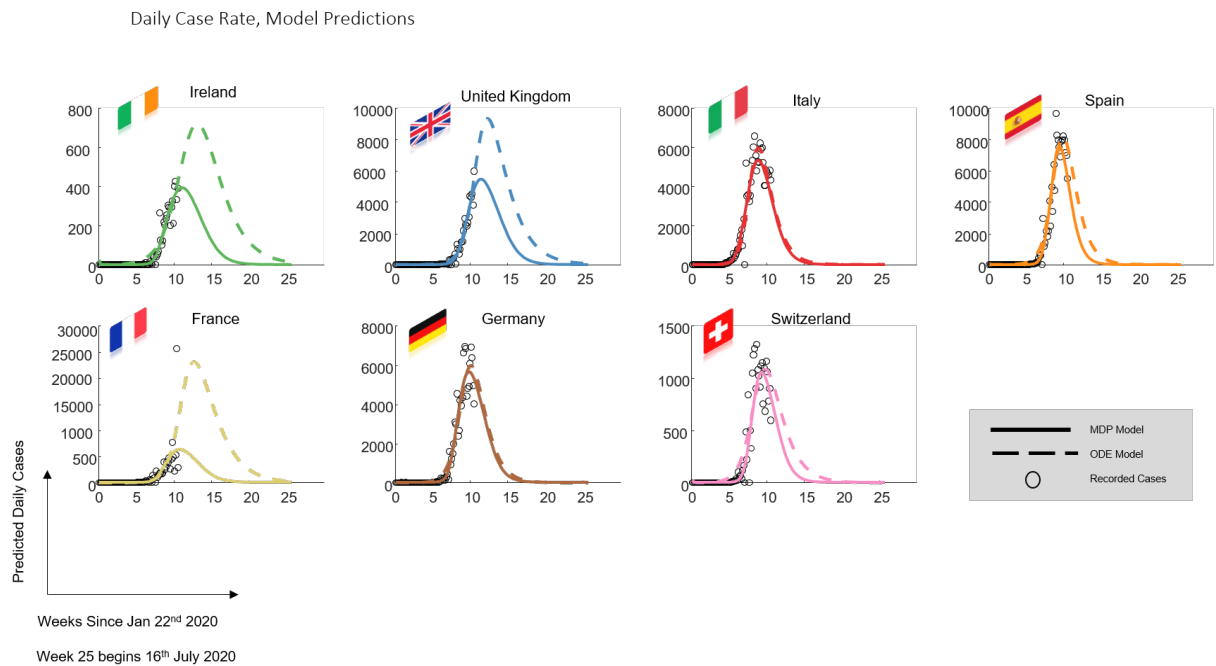

**Figure 3: Model predictions of daily *reported* case rates across countries. Note both models assume that many infected individuals will not be tested or reported in the daily case rates.**

The cumulative deaths (Figure 4) evinced relatively small discrepancies between the models, with the ODE model predicting a larger cumulative death toll, of 1250 for Ireland compared with 1008 deaths in Ireland given by the HMM model. For the UK the ODE and HMM were remarkably consistent, predicting a cumulative death toll of 49296 and 49785, respectively[1] (Figure 4). In other European countries, however, the discrepancies between the model predictions were greater in some countries such as France, Spain and Switzerland, with the HMM suggesting considerably lower cumulative deaths.

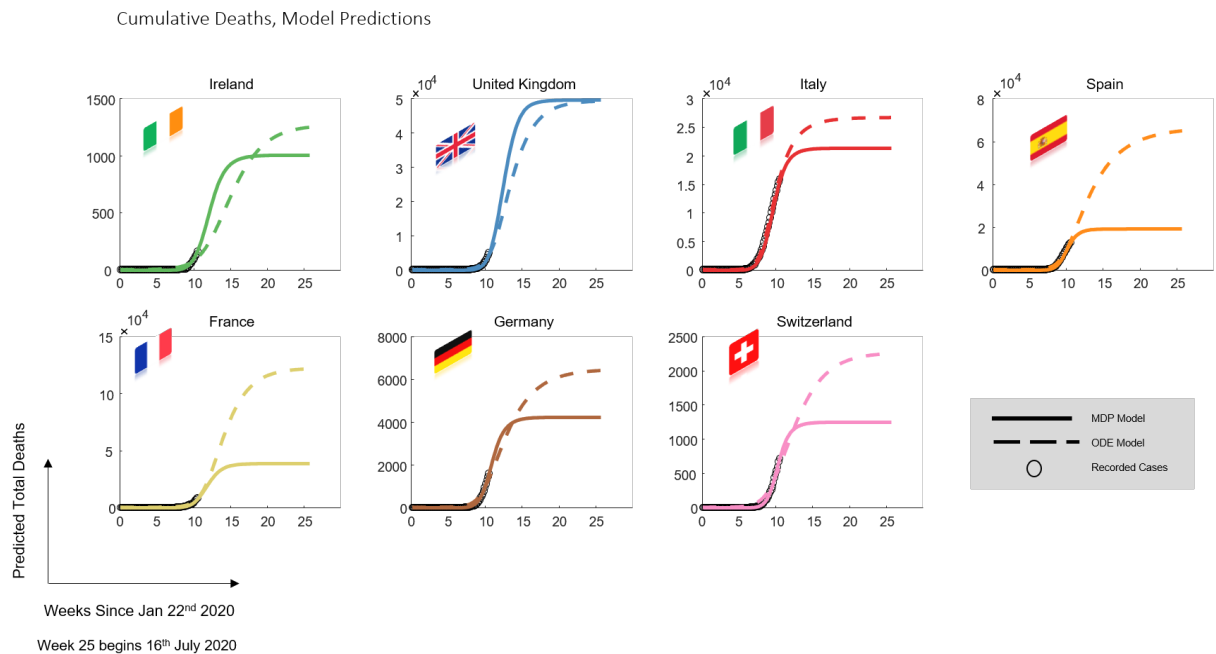

**Figure 4: Model predictions of cumulative deaths across countries.**

Finally, to test the assumption that low S0 proportions of the population may be indicative of a 'next wave' or several 'next waves', we estimated the initial susceptible size from the initial peak of the Spanish Flu pandemic of 1918-1919. Using data collated from approximately half of the United Kingdom: i.e. a population of approximately 22 million. Using the HMM model and variational Laplace, we see fits to the data that capture the falling peak. Here, we estimated the effective or susceptible population size was S0 = 4.03% of the total population size (Figure 5). Though dramatically different in terms of hospital care—the general picture remains - that large waves may be possible after low S0.

---

[1] These predictions fall to substantially lower levels, when empirical priors from a hierarchical or parametric empirical Bayes analysis that incorporates data from all countries are used.

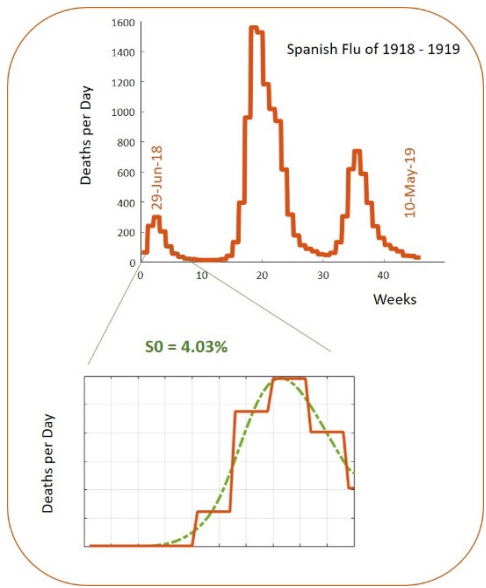

Figure 5. Spanish Flu Pandemic of 1918-1919 from regions of England and Wales. Initial S0 estimates from the first peak have a similar size (S0 Spanish Flu = 4.03%) to those estimated for the current coronavirus pandemic (S0 Corona = 6.4%).

Discussion

We used a variational Bayesian scheme (Friston, Mattout et al. 2007) to optimise the parameters of two distinctly constructed models of viral transmission (Friston, Parr et al. 2020, Moghadas, Shoukat et al. 2020). We optimised the parameters of these models based on daily reported cases and daily reports of death due to Covid-19. We optimised the model from data acquired for seven European countries. Both models were able to predict (i.e. fit) the current epidemic dynamics with plausible estimated trajectories. The models differed in their exact case rate predictions but predicted commensurate figures for the deaths in the United Kingdom and Ireland. How do these estimates relate to previous predictions of Covid-19 deaths in the UK? It was predicted (Ferguson, Laydon et al. 2020) that without interventions 510,000 deaths could occur, in the UK due to Covid-19. This analysis (Ferguson, Laydon et al. 2020) also predicted, that even with an optimal mitigation scenario, these death rates would reduce only by one half i.e. to 255,000. Thus, the predicted death cases of our models ~50,000 in the current cycle are in line with the predictions of mitigation effects, if we assume that several more cycles are possible.

Importantly, both models predicted that we are currently nearing or past the peak of daily case rates in all seven countries. However, the estimates suggest that after this cycle more than 80% of each country's total population in all countries studied remain susceptible. Therefore, we assume that future cycles will occur.

The predicted S0 was higher for the ODE model relative to the HMM model: in turn, the ODE model predicted a more prolonged cycle in the current period relative to the HMM model. This speaks to a trade-off between S0 and cycle times. Assuming herd immunity requires 60% of the susceptible population to be immune (Cohen and Kupferschmidt 2020), where 60%—one may conclude that further cycles are possible. However, that is not to say that populations within current outbreak areas may not reach herd immunity after the current cycle. Yet, if this is the case (immunity is clustered in geographic or some other organisation of communities), then parts of the country—particularly those communities with high contact numbers that have not 'been involved' in the current cycle may be more likely to participate in future cycles. And while it is obviously unrealistic to suppose that an additive linear effect of populations will emerge (Sirakoulis, Karafyllidis et al. 2000, Eubank, Guclu et al. 2004) (i.e., identically shaped cycles), given the complexity of contacts and population movement, our analysis may offer a rough guide to cycle immunity numbers.

As with most scientific research at this time, the modelling described above was conducted with haste. In line with the sentiments of the World Health Organization's Dr Mike Ryan *'Perfection is the enemy of the good when it comes to emergency management. Speed trumps perfection. And the problem in society we have at the moment is everyone is afraid of making a mistake. Everyone is afraid of the consequence of error, but the greatest error is not to move, the greatest error is to be paralyzed by the fear of failure.'* [1] Therefore, we are grateful to the coding repositories listed below, where interested researchers can reproduce or nuance our analyses.

**Code Availability**

For code and data see: https://github.com/RosalynMoran/Covid-19.git

[1] https://www.rev.com/blog/transcripts/world-health-organization-covid-19-update-march-13-2020

**Appendix A**

**ODE FLOW FUNCTION**

% States

```
S        = x(1); % susceptible
E        = x(2);% exposed
I_NH     = x(3); % infected will not require hospitalization
I_NH_SI  = x(4); % infected will not require hospitalization – socially isolated
I_RH     = x(5);  % infected will Require Hospitalization
I_RH_SI  = x(6);  % infected will Require Hospitalization – socially isolated
```

```
I_SC      = x(7);% infected asymotmatic - Sub Clinical
I_SC_SI   = x(8);% infected asymotmatic - Sub Clinical – socially isolated
I_H       = x(9); % Hospitalized NOT n ICU
I_ICU     = x(10); % Hospitalized in the ICU
R         = x(11); % recovered
D         = x(12); % deaths

% Parameters

N0      =  (exp(P.N))*1e6;
beta    = (0.09*exp(P.beta))*(10*exp(P.k));      %  beta x contacts
betaSI  = (0.09*exp(P.beta))*(2*exp(P.k_SI));    % contacts in isolation
gamma   = (1/4.6)*exp(P.gamma);                  % recovery rate = 1/days_infection 4.6 from pnas, 3 or 7 from lancet

kappa   = (1/5.2)*exp(P.kappa); % latency rate 1/(days incubation) - assume 5.2 days
alpha   = (1/2)*exp(P.alpha);   % reduced transmission rate for subclinical cases - a prior one quarter PNAS - one half
alpha   = min([alpha 0.5]);

p_clin  = (1/4)*exp(P.p);     % proportion of cases that are symptomatic - that may or may not need hospitalization
p_clin  = min([p_clin 0.5]);

delta   = (1/7)*exp(P.delta);  % 1/average time from symptom onset to hospitalization

h       =  (1/2)*exp(P.h) ;% proportion of clinical cases requiring hospital or ICU
h       =  min([h 0.5]);

c = (1/10)*exp(P.c);   % proportion in hospital requiring ICU- estimate 10% -PNAS 1/100
c = min([c 0.75]);

mh = 0.02*exp(P.mh); % weight of death rate among non ICU
mh = min([mh 0.5]);

muh  = (1/9.7)*exp(P.muh)  ;  % 1/average days from hospitalized non-ICU deaths
phih = (1/10)*exp(P.phih); % length of hospital stay before recovery

mc =  0.13*exp(P.mc) ; % weight of death rate among  ICU
mc =  min([mc 0.75]);

muc =  (1/7)*exp(P.muc)  ; % 1/average days from hospitalized ICU to deaths
phic = (1/14)*exp(P.phic) ;  % length of ICU stay before reocvery

q  = 0.05*exp(P.q);
q  = min([q 1]);

fi = 0.8*exp(P.fi);
fi = min([fi 1]);

taui = 1*exp(P.taui);
fa   = 0.05*exp(P.fa);
fa   = min([fa 1]);

taua =  (1/2)*exp(P.taua);

%- Dynamics

dS_dt    = -beta*(S/N0)*(I_NH + I_RH) - alpha*beta*(S/N0)*I_SC...
   -betaSI*(S/N0)*(I_NH_SI + I_RH_SI) - alpha*betaSI*(S/N0)*I_SC_SI;

dE_dt    = beta*(S/N0)*(I_NH + I_RH)  + alpha*beta*(S/N0)*I_SC ...
   + beta*(S/N0)*(I_NH_SI + I_RH_SI) + alpha*beta*(S/N0)*I_SC_SI- kappa*E;
```

```
dI_NH_dt    = p_clin*(1-q)*(1-h)*kappa*E  -  (1-fi)*gamma*I_NH - (fi)*taui*I_NH;
dI_NH_SI_dt = p_clin*q*(1-h)*kappa*E   -  gamma*I_NH_SI + (fi)*taui*I_NH;

dI_RH_dt    = p_clin*(1-q)*(h)*kappa*E   -  (1-fi)*delta*I_RH - fi*taui*I_RH;
dI_RH_SI_dt = p_clin*q*(h)*kappa*E    -  delta*I_RH_SI + fi*taui*I_RH;

dI_SC_dt    = (1-p_clin)*kappa*E     - (1-fa)*gamma*I_SC - (fa)*taua*I_SC;
dI_SC_SI_dt = (fa)*taua*I_SC    - gamma*I_SC_SI;

dH_dt    = (1-c)*(1-fi)*delta*I_RH +  (1-c)*delta*I_RH_SI  - (mh*muh + (1-mh)*phih)*I_H;   %
dICU_dt  = c*(1-fi)*delta*I_RH + c*delta*I_RH_SI     - (mc*muc + (1-mc)*phic)*I_ICU; %

dR_dt   = gamma*(I_SC + I_NH + I_SC_SI + I_NH_SI) +  (1-mh)*phih*I_H  + (1-mc)*phic*I_ICU;
dD_dt   = mh*muh*I_H + mc*muc*I_ICU ;
```

%%

**ODE Observer FUNCTION**

```
cases =   prop_asymp*(x(7) +x(8)) +  prop_sympNH*(x(3)+x(4)) +  x(5) + x(6); % twenty percent of Symptomatic not hospitalised cases tested one percent of Asymptomatic tested

deaths =  x(12);

prop_asymp  = 0.01*exp(P.cases_from_SC);
prop_sympNH = 0.2*exp(P.cases_from_NH);
prop_asymp  = min([prop_asymp, 1]);
prop_sympNH = min([prop_sympNH, 1]);
```

**ODE CITY Observer FUNCTION**

```
cases =  P.no_cities*(prop_asymp*(x(7) +x(8)) +  prop_sympNH*(x(3)+x(4)) +  x(5) + x(6)); % twenty percent of Symptomatic not hospitalised cases tested one percent of Asymptomatic tested

deaths =  P.no_cities*x(12);
```